\documentclass[journal=jacsat,manuscript=article]{achemso}

\usepackage[version=3]{mhchem} 
\usepackage[utf8]{inputenc}
\usepackage{babel}
\usepackage{makecell}
\usepackage{comment}

\author{Justin Boddison-Chouinard}
    \affiliation{Department of Physics, University of Ottawa, Ottawa, Ontario, K1N 9A7}
\author{Alex Bogan}
    \affiliation{Emerging Technologies Division, National Research Council of Canada, Ottawa, Ontario, K1A 0R6}
\author{Norman Fong}
    \affiliation{Emerging Technologies Division, National Research Council of Canada, Ottawa, Ontario, K1A 0R6}
\author{Pedro Barrios}
    \affiliation{Advanced Electronics and Photonics, National Research Council of Canada, Ottawa, Ontario, K1A 0R6}
\author{Jean Lapointe}
    \affiliation{Advanced Electronics and Photonics, National Research Council of Canada, Ottawa, Ontario, K1A 0R6}
\author{Kenji Watanabe}
    \affiliation{Research Center for Functional Materials, National Institute for Materials Science, 1-1 Namiki, Tsukuba 305-0044, Japan}
\author{Takashi Taniguchi}
    \affiliation{International Center for Materials Nanoarchitectonics, National Institute for Materials Science,  1-1 Namiki, Tsukuba 305-0044, Japan}
\author{Adina Luican-Mayer}
    \affiliation{Department of Physics, University of Ottawa, Ottawa, Ontario, K1N 9A7}
    \email{luican-mayer@uottawa.ca}
\author{Louis Gaudreau}
    \email{Louis.Gaudreau@nrc-cnrc.gc.ca}
    \affiliation{Emerging Technologies Division, National Research Council of Canada, Ottawa, Ontario, K1A 0R6}

\title{Charge detection using a WSe$_2$ van der Waals heterostructure}

\begin{document}

\begin{abstract}
    Detecting single charging events in quantum devices is an important step towards realizing practical quantum circuits for quantum information processing. In this work, we demonstrate that van der Waals heterostructure devices with gated nano-constrictions in monolayer WSe$_2$ can be used as charge detectors for nearby quantum dots. These results open the possibility of implementing charge detection schemes based on 2D materials in complex quantum circuits. 
 
\end{abstract}


The detection and manipulation of individual charges and spins is at the heart of developing quantum technologies. In particular, the detection of individual charges in a  quantum dot (QD) can be realized by placing a quantum point contact (QPC) in its vicinity and detecting changes in the conductance through the QPC current as a result of the altered electrostatic environment around the QD. While transport directly through a single quantum dot and charged detection are equivalent this is not the case for circuits containing more than a single quantum dot. In that situation direct transport only occurs under specific resonance conditions between dots which becomes severely restrictive as additional quantum dots are added. This is in contrast to charge detection techniques which are able to identify the charge in each dot separately without current flowing directly through the circuit. As a result charge detection technology is a requirement for complex spin qubit quantum circuits\cite{Elzerman2004, Petta2005, Granger2012}. The implementation of this approach was initially successfully demonstrated in GaAs two-dimensional electron gas (2DEG) devices\cite{Hasko1993}. More recently, two-dimensional (2D) materials including graphene and semiconducting transition metal dichalcogenides (TMDs) emerged as a novel platform to realize electrostatically confined quantum circuits. Within the 2D materials platform, QPC's have been demonstrated in TMDs\cite{Pisoni2017, Zhang2017, Marinov2017, Wang2018, Epping2018,Sakanashi2021}, and more specifically, QPC charge detection has been demonstrated in monolayer and bilayer graphene (BLG) devices \cite{Ensslin2008, GuoGP2010, Ihn2011, Stampfer2013, Stampfer2011, Ensslin2019}. However, charge detection has not yet been implemented in semiconducting TMDs. The direct band gap in TMDs, along with the interplay of the spin and valley degrees of freedom, makes them an appealing platform for realizing hybrid quantum circuits with optical and solid state qubits\cite{Fujita2019, Kosaka2012, Gaudreau2017}. Additionally, the strong spin-orbit coupling in TMDs could enable fast spin qubit manipulation with AC electric fields\cite{Nowack2007} and lead to the spin-valley locking, by which qubits could be more robust against electro-magnetic noise.  Although the first gated quantum dots in TMDs have been recently realized\cite{Pisoni2017, Zhang2017, Pisoni2018, Wang2018, Davari2020, Boddison2021}, charge detection using gated nano-constrictions is yet to be implemented. In this work, we realize a nano-constriction in a monolayer WSe$_2$ channel and, demonstrate for the first time, charge detection of a gate controlled WSe$_2$ quantum dot.

The charge detector device used in this work is based on monolayer $\mathrm{WSe_2}$, schematically described in Figure 1a. Its optical micrograph is shown in Figure 1b. The device was assembled on a $\mathrm{ p-Si/SiO_2}$ (285 nm) substrate using standard dry transfer methods. Local nano-constriction gates [Ti (2.5 nm) / Au (2.5 nm)] were first deposited on the substrate using electron beam lithography (Fig 1b inset). They are 100nm in width, with a lateral spacing of 50 nm, and varying gap spacing (50 nm, 100 nm, and 75 nm). The local nano-constriction gates are responsible for locally depleting the carriers in the monolayer $\mathrm{WSe_2}$ flake. A flake of hexagonal boron nitride (hBN) with 44nm thickness was transferred on top of the pre-patterned gates and electrical contacts [Cr (2 nm) / Pt (8 nm)] were subsequently lithographically patterned on top of the hBN. To increase sample cleanliness and, consequently, reduce contact resistance, the device was cleaned in a vacuum furnace ($10^{-7}$  Torr) at 300$^{\circ}$C for 30 minutes. Additionally, remaining polymer residues were removed using an atomic force microscope (AFM) tip in contact mode\cite{Goossens2012, Rosenberger2018, Boddison2021}. A $\mathrm{WSe_2}$ monolayer and a 37 nm thick hBN flake were exfoliated and identified on independent substrates. Using dry transfer techniques, the hBN flake was first picked-up an then used to pick-up the $\mathrm{WSe_2}$ flake; subsequently, the $\mathrm{WSe_2}$/hBN stack was precisely placed on top of the cleaned electrical contacts. We note that prior to the deposition of the hBN/$\mathrm{WSe_2}$ stack, the $\mathrm{WSe_2}$ surface was also mechanically cleaned using an AFM tip. A last lithographic step was performed to deposit a top gate [Ti (5 nm) / Pd (20 nm) / Au (100 nm)] used to globally modify the charge carrier density of the $\mathrm{WSe_2}$ flake and to achieve ohmic contacts.

\begin{figure}[t]
    \centering
    \includegraphics[width=\textwidth]{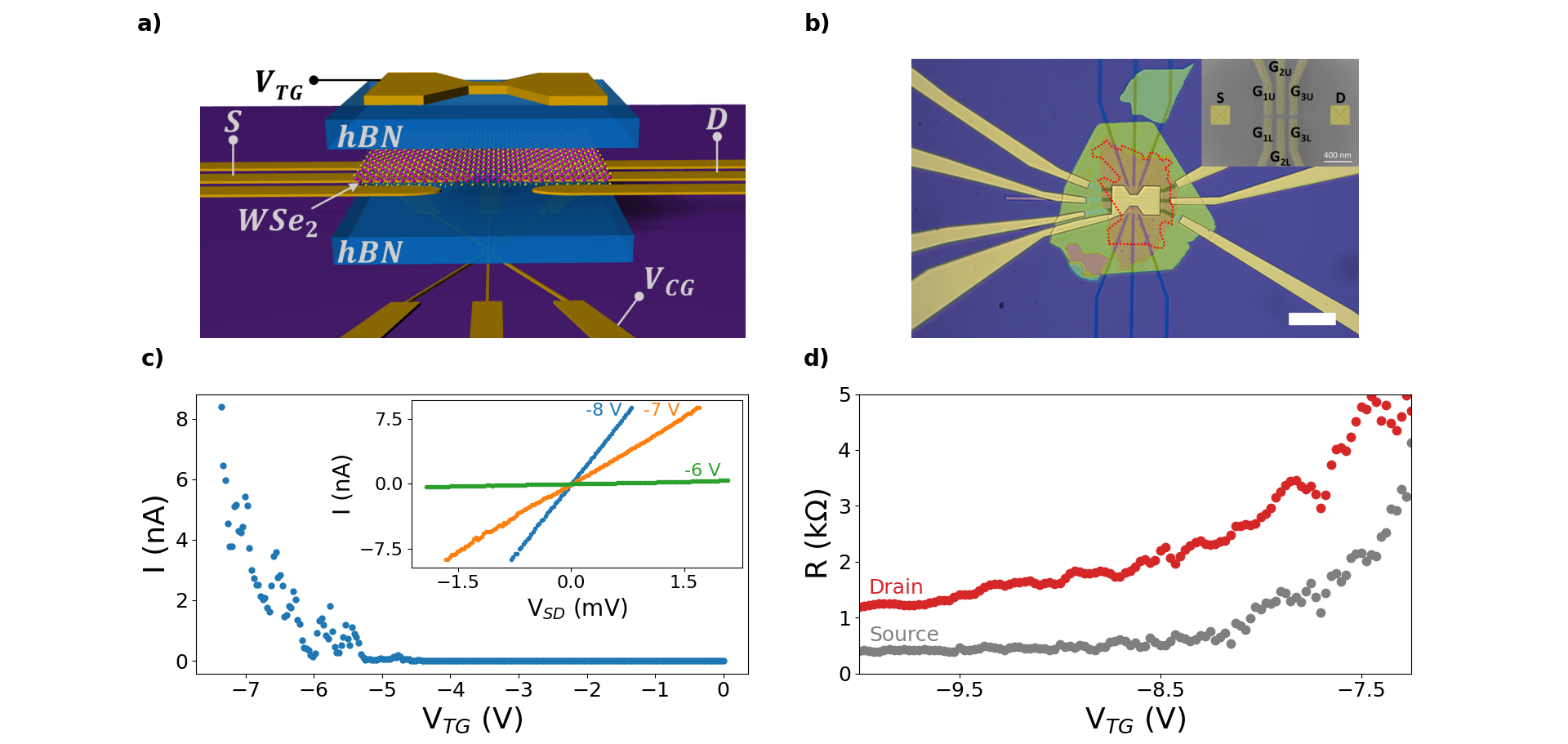}
    \caption{(a) Schematic architecture of the charge detection device based on monolayer $\mathrm{WSe_2}$. (b) Optical micrograph of the completed device used in this work. The $\mathrm{WSe_2}$ flake is outlined in red for clarity. Scale bar: $20 ~\mu m $. Inset: false-color scanning electron micrograph of the control gates. (c) Source-drain current as a function of the top gate voltage ($\mathrm{V_{TG}}$) showing the activation curve of the device. Inset: linear I-V characteristic observed when $\mathrm{V_{TG} < - 6 ~V}$. The temperature of these measurements was $\mathrm{T = 4 ~K}$. (d) Contact resistance of the source and drain contacts as a function of the top gate voltage $\mathrm{V_{TG}}$, as indicated. The grey (red) curve represents the resistance associated to the source (drain) contact. The temperature of these measurements was $\mathrm{T = 1.5 ~K}$.}
    \label{fig:fig1}
\end{figure}

In a first measurement step, the source-drain current was monitored as a function of the top gate voltage ($\mathrm{V_{TG}}$) at a temperature of $\mathrm{T = 4 ~K}$ as shown in Figure 1c. The current increases at higher negative values of $\mathrm{V_{TG}}$, indicating hole conduction across the $\mathrm{WSe_2}$ channel. At top gate voltage values below – 6 V, the current voltage characteristics across the source-drain contacts show a linear behaviour, as presented  for select values of ($\mathrm{V_{TG}}$) in the inset of Figure 1c. One of the principal challenges in developing quantum devices in 2D semiconductors is realizing good electrical contacts. Thus, when the temperature was further reduced ($\mathrm{T = 1.2 ~K}$), by using a three-terminal measurement scheme, the contact resistance of the source and drain were individually measured as a function of ($\mathrm{V_{TG}}$) as seen in Figure 1d. The importance of the top-gate for achieving high quality contacts is evidenced by the low contact resistance measured when $\mathrm{V_{TG} = - 10 ~V}$. At this top-gate voltage, a contact resistance of 404 ${\Omega}$ was measured for the source contact and 1240 ${\Omega}$ for the drain contact.



We then investigated the influence of the local control gates on the transport properties of the device in a dilution refrigerator with a base temperature below 10 mK. Specifically, the pair of gates $\mathrm{G_{3L}}$ and $\mathrm{G_{3U}}$ (right-most control gates in the inset of Figure 1b) were studied while all other control gates were grounded (Figure 2). A bias voltage of $\mathrm{500 \mu V}$ was applied to the source contact, while the drain contact was grounded. A gradually increasing voltage was applied to both $\mathrm{G_{3L}}$ and $\mathrm{G_{3U}}$ to create a channel by locally depleting the $\mathrm{WSe_2}$ flake of holes in the areas directly above the two control gates. Keeping $\mathrm{G_{3L}}$ at $\mathrm{15.12 ~V}$, $\mathrm{G_{3U}}$ was varied, while the source-drain current $I$ was monitored, as shown in Figure 2a (blue dotted curve). The source-drain current decreased as a function of $\mathrm{G_{3U}}$ until $\mathrm{G_{3U} = 15.6 ~V}$, when it suddenly increased, and, as $\mathrm{G_{3U}}$ is further increased, the current eventually pinches off. We demonstrate below that it is related to the nano-constriction acting as a non-invasive probe of the charge of a nearby incidental quantum dot.

Figure 2b illustrates the model for the blue current trace characteristics observed in Figure 2a. A nearby quantum dot has an initial population of $\mathrm{N}$ holes. As a larger positive voltage is applied to the local control gate $\mathrm{G_{3U}}$, the width of the channel formed between $\mathrm{G_{3U}}$ and $\mathrm{G_{3L}}$ decreases, resulting in a decrease of current. At the same time, $\mathrm{G_{3U}}$ controls the chemical potential of the nearby quantum dot and can therefore remove holes from it by applying a larger positive voltage. At a specific gate voltage, the population of the quantum dot transitions from $\mathrm{N}$ holes to $\mathrm{N-1}$ holes, resulting in a widening of the channel and, as a result, a sudden increase in current through the charge detector.

 Assuming no change in the number of holes (N) in the quantum dot as a function of gate voltage, the current is expected to follow the leftmost grey dashed curve in Figure 2a. If a hole is removed from the quantum dot, which would be therefore populated by N-1 holes, a larger voltage applied to $\mathrm{G_{3U}}$ would be needed to achieve the same result (rightmost grey dashed curve). Thus, the jump between the two gray curves indicates a switch between a nearby quantum dot population of N and N-1 holes.

\begin{figure}[t]
    \centering
    \includegraphics[width=\textwidth]{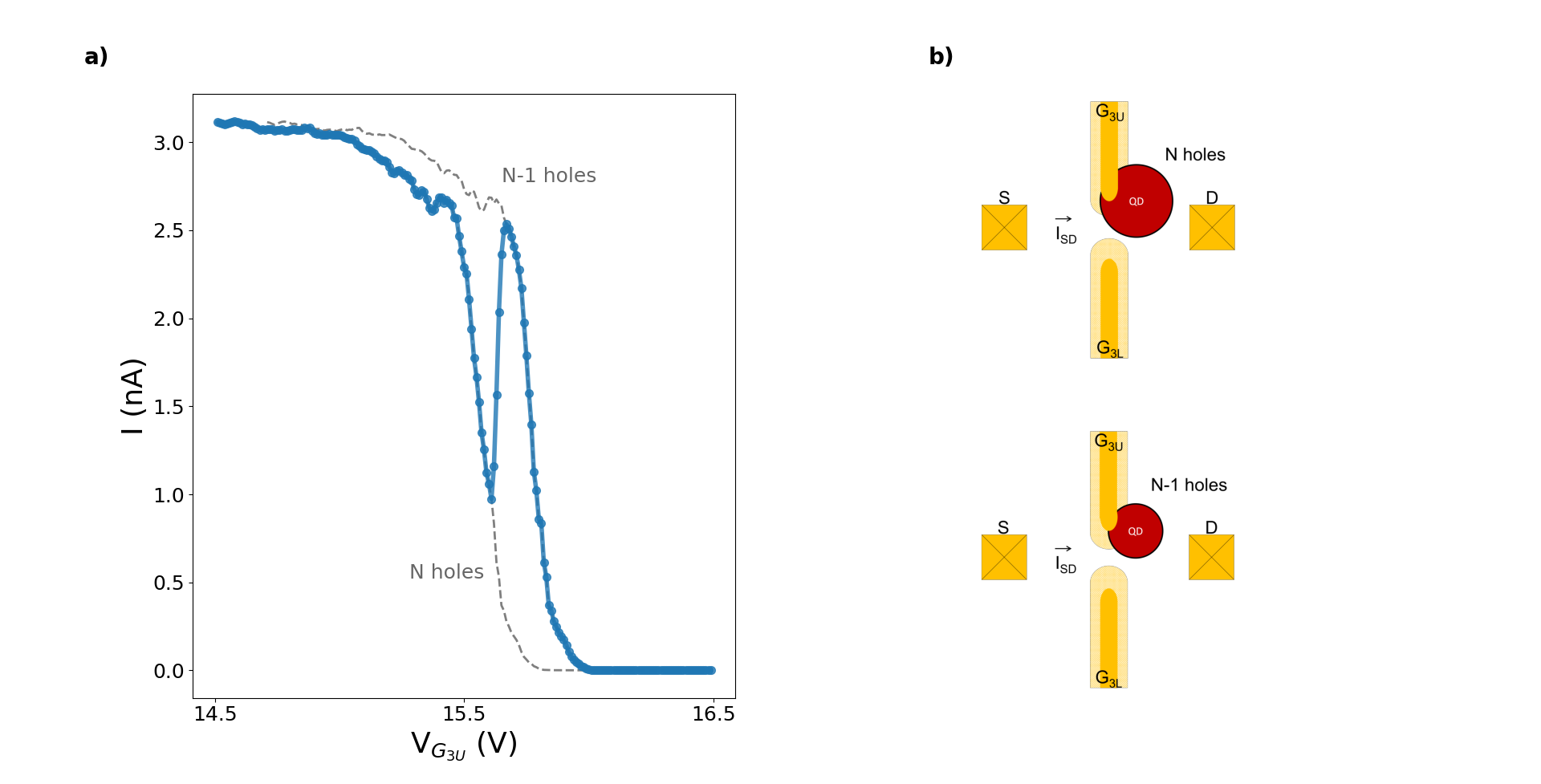}
    \caption{(a) The blue dotted curve is the source-drain current as a function of the local control gate $\mathrm{G_{3U}}$, while the local control gate $\mathrm{G_{3L} = 15.12 ~V}$. The grey dashed curves are guides for two distinct quantum dot populations N and N-1, as indicated. (b) Schematic explanation of the result observed in (a), where a change in the hole population of a nearby quantum dot affects the channel width and the source-drain current. $\mathrm{T < 7 ~mK}$.}
    \label{fig:fig2}
\end{figure}



To demonstrate that the observed step in the source-drain current of Figure 2 is caused by a change in the charge population of a nearby quantum dot, we studied the evolution of the charge detection step as a function of $\mathrm{G_{3U}}$ for different values of $\mathrm{G_{3L}}$ between 15 V (dark red curve) and 15.4 V (dark blue curve) as shown in Figure 3a. We observe that as $\mathrm{G_{3L}}$ increases, the constriction pinch-off voltage decreases as a function of $\mathrm{G_{3U}}$, while the voltage at which the charge transfer step occurs remains mostly  constant. This indicates that the origin of the step is distinct from the nano-constriction. In addition, we note that a current peak is still visible, but with considerably less amplitude, after the detector is pinched-off for higher $\mathrm{G_{3L}}$ values. We attribute this behaviour to parallel transport through the quantum dot in the Coulomb Blockade regime.

To investigate this further, we extract the position of three distinct events in each current trace of Figure 3a: the first pinch-off, the second pinch-off and the charge transfer step. The first pinch-off point is extracted by taking a linear regression of the initial downward slope of the current and obtaining its x-intercept. Similarly, the second pinch-off point is the x-intercept of the linear regression which fits the current’s second downward slope. The charge transfer step is taken to be the middle point between the current’s local minimum and local maximum around the step. An example of this is illustrated in Figure 3b. This analysis is performed for all values of $\mathrm{G_{3L}}$ and plotted in Figure 3c. The slope associated to the first pinch-off point and the charge transfer step are each constant as a function of $\mathrm{G_{3L}}$ and have values of $\mathrm{-1.7 \pm 0.1 ~V_{G_{3U}}/ V_{G_{3L}}}$ and $\mathrm{-0.34 \pm 0.03 ~V_{G_{3U}}/ V_{G_{3L}}}$ respectively. The difference in slopes between the linear dependencies of these events (first pinch-off, and charge transfer) further indicate that the origin of the charge transfer step and the nano-constriction are independent.

Looking at the extrapolations associated to the second pinch-off point, we observe a change in slope at $\mathrm{V_{G_{3L}} = 15.22 ~V}$, the voltage where the charge transfer step transitions into a Coulomb blockade peak, associated to parallel transport through the quantum dot.  The slope corresponding to the second pinch-off point in the charge detection regime ($\mathrm{-1.6 \pm 0.2}$ \linebreak
$\mathrm{~V_{G_{3U}}/ V_{G_{3L}}}$) is similar to the slope associated to the first pinch-off point, suggesting its origin being the nano-constriction. However, in the Coulomb blockade regime, the slope corresponding to second pinch-off ($\mathrm{-0.35 \pm 0.05 ~V_{G_{3U}}/ V_{G_{3L}}}$) no longer follows the first pinch-off, but instead it follows the charge transfer step. This is due to the fact that at $\mathrm{V_{G_{3L}} > 15.22 ~V}$ the constriction is completely pinched off and we are measuring only parallel transport through the quantum dot. We note that the quantum dot transport is still present in the charge detection regime, however the signal is dominated by the transport through the nano-constriction.

\begin{figure}[t]
    \centering
    \includegraphics[width=\textwidth]{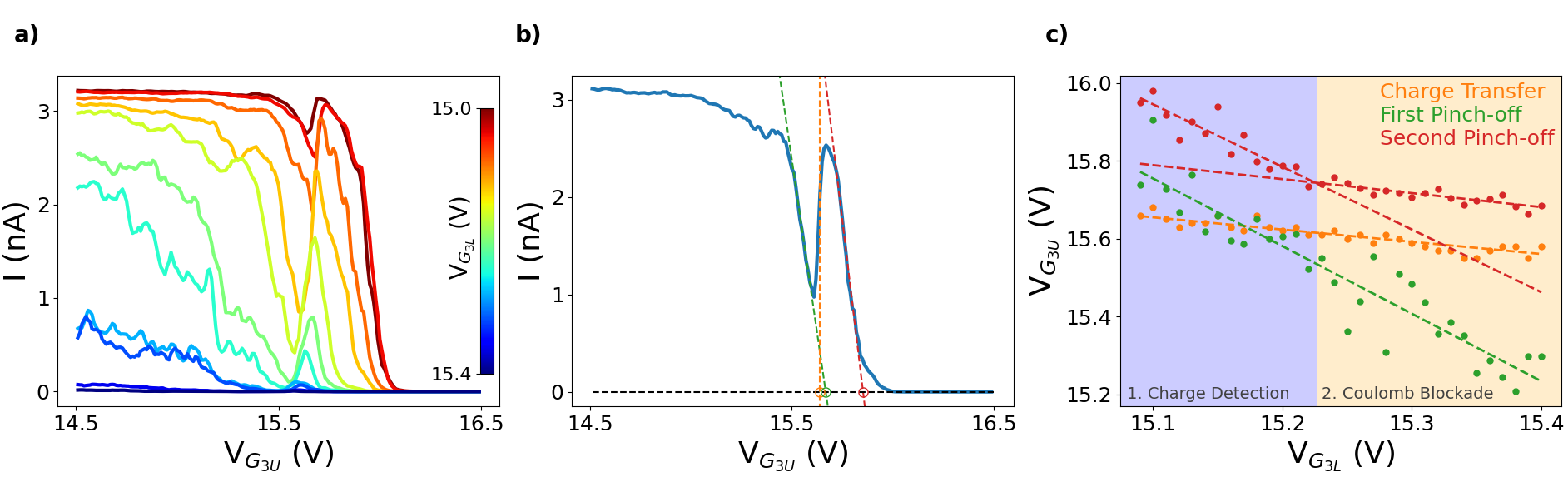}
    \caption{(a) Source-drain current as a function of the local control gate $\mathrm{G_{3U}}$. Curves are taken for different values of $\mathrm{G_{3L}}$ ranging from 15 V (dark red) to 15.4 V (dark blue). Experimentally, curves were taken at an interval of $\mathrm{{\Delta}G_{3L} = 0.01 ~V}$ but for clarity, an interval of $\mathrm{{\Delta}G_{3L} = 0.05 ~V}$ is shown here. (b) An example of how the “position” of the first pinch-off (green), the second pinch-off (red) and the charge transfer step (orange) are obtained. The trace is taken when the voltage on  $\mathrm{G_{3L}} = 15.12$. (c) The position of the first pinch-off (green), the second pinch-off (red) and the charge transfer step (orange) plotted as a function of $\mathrm{G_{3L}}$ along with fitted linear regression. $\mathrm{T < 7 ~mK}$. }
    \label{fig:fig3}
\end{figure}



\begin{figure}[t]
\begin{minipage}{\textwidth}
    \begin{minipage}[b]{0.48\textwidth}
        \centering
        \includegraphics[width=\textwidth]{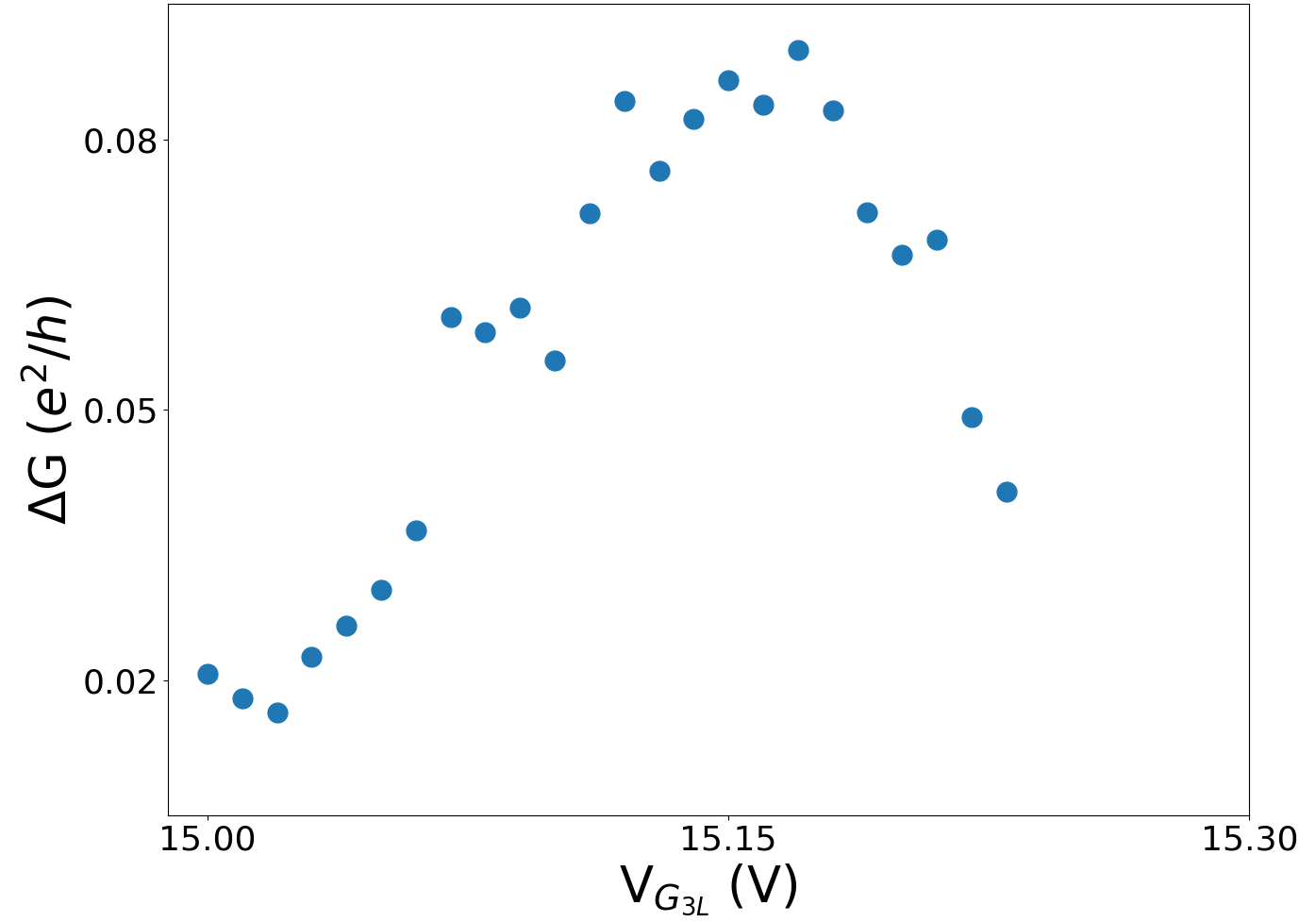}
        \captionof{figure}{The charge detection sensitivity ($\mathrm{{\Delta}G}$) of the measured device as a function of the voltage applied to $\mathrm{G_{3U}}$.}
    \end{minipage}
    \hfill
    \begin{minipage}[b]{0.48\textwidth}
        \centering
        \begin{tabular}{ccc}\hline
            Reference & $\mathrm{{\Delta}G} ~(e^2/h)$ & Material\\ \hline
            \thead{Appl. Phys. Lett. \\ (\citeyear{Ensslin2008})\cite{Ensslin2008}}  & $10^{-4}$  & Graphene\\
            \thead{Appl. Phys. Lett. \\ (\citeyear{GuoGP2010})\cite{GuoGP2010}} & $<0.1$  & Graphene\\
            \thead{Phys. Status Solidi B \\ (\citeyear{Stampfer2011})\cite{Stampfer2011}} & $0.005$ & BLG\\
            \thead{Phys. Rev. B \\ (\citeyear{Ihn2011})\cite{Ihn2011}} & $0.02$ & Graphene\\
            \thead{Nat. Commun. \\ (\citeyear{Stampfer2013})\cite{Stampfer2013}} & $0.002$  & Graphene\\
            \thead{Nano Lett. \\ (\citeyear{Ensslin2019})\cite{Ensslin2019}} & $0.2$  & BLG\\
            \thead{This work \\ (2022)} & $0.09$ & ML WSe$_2$\\ \hline
        \end{tabular}
        \captionof{table}{Comparison of the charge detector sensitivity ($\mathrm{{\Delta}G}$) in various 2D material devices.}
    \end{minipage}
\end{minipage}
\end{figure}

The sensitivity of the charge detector, taken as the change in conductance ($\mathrm{{\Delta}G}$) per hole in units of $\mathrm{e^2/h}$, was tuned as a function of the voltage applied to $\mathrm{G_{3L}}$ as can be seen in figure 4a. An optimal value of $\mathrm{0.09 e^2/h}$ is reached when $\mathrm{G_{3L} = 15.17 ~V}$. Table 1 compares this sensitivity with previously reported charge detectors using graphene and bilayer graphene devices, suggesting that TMDs show promising prospects of being used in charge detection scheme based on nano-constrictions.


In summary, we fabricated van der Waals hetetostructure devices where we electrostatically define nano-constrictions in monolayer WSe$_2$. Our fabrication technique results in very low contact resistance, which allows the study of transport properties through the constrictions. By monitoring the current in a nano-contriction, we demonstrate for the first time using TMDs, the possibility of using such a device as a charge sensor for nearby quantum dots. The sensitivity of the WSe$_2$ is comparable to similar devices implemented in graphene and bilayer graphene. These results pave the way for future the development of complex quantum circuits, where charge detectors can be integrated into gated quantum dot architectures.

\begin{acknowledgement}

This work was supported by the High Throughput and Secure Networks Challenge Program at the National Research Council of Canada.
AL-M and JBC acknowledge funding from the National Sciences and Engineering Research Council (NSERC) Discovery Grant RGPIN-2016-06717.  We also acknowledge the support of the Natural Sciences and Engineering Research Council of Canada (NSERC) through Strategic Project STPGP 521420.

K.W. and T.T. acknowledge support from JSPS KAKENHI (Grant Numbers 19H05790, 20H00354 and 21H05233).

\end{acknowledgement}

\bibliography{main}

\end{document}